\documentclass[twocolumn,showpacs,superscriptaddress,preprintnumbers,amsmath,amssymb]{revtex4}

\usepackage{color}
\usepackage{graphicx}
\usepackage{bm}

\begin{document}

\title{Magnetic moment formation at a dilute $^{140}$Ce impurity in $R$Co$_2$ compounds}

\author{A.L. de Oliveira}
\email{alexandre.oliveira@ifrj.edu.br}
\affiliation{Instituto Federal de Educa\c{c}\~{a}o, Ci\^{e}ncia e Tecnologia do Rio de  Janeiro, Rua L\'{u}cio Tavares 1045, Nil\'{o}polis, 26530-060, RJ, Brazil.}

\author{C.M. Chaves}

\affiliation{Centro Brasileiro de Pesquisas F\'{\i}sicas, Rua Dr. Xavier Sigaud 150, 22290-180, Rio de Janeiro, Brazil}

\author{N.A. de Oliveira}
 \affiliation{Instituto de F\'{\i}sica, Universidade do Estado do Rio de Janeiro, Rua S\~{a}o Francisco Xavier 524, 20550-013, Rio~de~Janeiro, RJ, Brazil}

\author{A. Troper}
\affiliation{Centro Brasileiro de Pesquisas F\'{\i}sicas, Rua Dr. Xavier Sigaud 150, 22290-180, Rio de Janeiro, Brazil}


\begin{abstract}
A great deal of experimental work using perturbed angular correlation (PAC) has succeed in measuring hyperfine fields in Ce diluted in metallic systems, thus allowing the determination of the local impurity moment at low temperatures. Motivated by such experimental work on $^{140}$Ce  placed on a $R$ site of the rare earth ($R$ = Gd, Tb, Dy, Ho, Er) in $R$Co$_{2}$,  we theoretically discuss, within a simple model, the local magnetic moments and thereby calculate the magnetic hyperfine fields. The results are in good agreement with the experimental data. For the sake of comparison we recall our previous results on Ta d-impurity embedded in the same hosts.
\end{abstract}

\maketitle

The description of the formation of local magnetic moments at impurities embedded in metallic systems has been the concern of condensed matter theorists since the pioneer work of Friedel.\cite{Friedel1958}
On the other hand, from the experimental point of view, the technic of time differential $\gamma-\gamma$ angular correlation has been largely used in the last years to obtain new information on the subject.\cite{saito} In this work we discuss the formation of magnetic moments theoretically in connection with magnetic hyperfine fields.

A single Ce impurity in $R$ site of $R$Co$_{2}$ Laves phase compounds introduces in the host the following main effects:
\begin{enumerate}
  
\item[(i)] 	a local potential due to the different impurity and host charges;

\item[(ii)] 	a local change in the Coulomb interaction $U^{I} \neq U^{R}$, the superscript $I$ referring to the impurity and $R$ to the rare earth ;

\item[(iii)] 	the hopping among neighboring sites is modified when the impurity is one of them.
 
\item[(iv)] Ce is in an intermediate valence state as inferred from experiments.\cite{leal87,linus} So, the Ce \textit{4f} resonance is fractionally occupied and is strongly correlated and admixed with the \textit{d}-host conduction state lying very close to the Fermi energy. 

\end{enumerate}

The Ce valence state ($4f^15d^16s^2$ in the atomic configuration) is $4f^{1-\delta}5d^{1+\delta}6s^2$, where $\delta$ is the amount of charge transfered from the 4f-resonance to the d-conduction band. It is claimed in the literature \cite{linus,ra} that the Ce valence state lies in the interval ($3.2$--$3.4$). In a previous calculation\cite{leal87} intended for the Ce\underline{Gd} we have self-consistently obtained $\delta=0.25$. In the present paper we adopted the same value.

The main ingredients of the intermetallic host are the two conduction bands (\textit{s-p} and \textit{d}, the last including Coulomb correlation) of the rare earth $R$, whose centers depend on the \textit{4f}-polarization of $R$. This is the way the  \textit{4f} electrons of the host rare earth $R$ enter into the model.

  The model includes also a magnetic coupling among the impurity and the Co magnetic moments.\cite{alo2}
The Hamiltonian describing the \textit{d} and \textit{s-p} contributions to the formation of the local magnetic moment is
\begin{equation}
\mathcal{H}=H_{R}+V+H_{M},
\label{eq:ham1}
\end{equation}
where
\begin{eqnarray}
H_{R} &=& \sum_{i,\sigma }\varepsilon _{\sigma }^{c{\rm h}}c_{i\sigma}^{\dagger}c_{i\sigma }
+\sum_{i,j,\sigma }t_{ij}^{c}c_{i\sigma }^{\dagger}c_{j\sigma} 
+\sum_{i,\sigma }\varepsilon _{\sigma }^{d{\rm h}}d_{i\sigma}^{\dagger}d_{i\sigma }
\nonumber \\
&&+\sum_{i, j,\sigma }t_{ij}^{d}d_{i\sigma }^{\dagger}d_{j\sigma} 
+\sum_{i}U^{R}n^{d}_{i\downarrow }n^{d}_{i\uparrow },
\label{eq:HR}
\end{eqnarray}
defines an effective pure rare earth host which consists of conduction \textit{s-p} and \textit{d} bands polarized by the \textit{4f} electrons.  In Eq.~(\ref{eq:HR}), $\varepsilon _{\sigma }^{c{\rm h}}$ and $\varepsilon _{\sigma }^{d{\rm h}}$ are the center of the {\it s-p} and \textit{d} energy bands, now depending on the spin $\sigma $ orientation, $c_{i\sigma}^{\dagger}$ ($c_{i\sigma}$) and $d_{i\sigma}^{\dagger}$ ($d_{i\sigma}$) are the creation (annihilation) operators of  electrons at site $i$ with spin $\sigma $ ($i=0$ is the impurity site) and $t^{c}_{ij}$ and $t^{d}_{ij}$ are the electron hopping energies between neighboring $i$ and $j$ sites.
The second term of Eq.~(\ref{eq:ham1}) is the non-local potential given by
\begin{eqnarray}
V&=&\sum_{\sigma}({\varepsilon _{\sigma }^{c{\rm I}}-\varepsilon _{\sigma }^{c{\rm h}})c_{0\sigma }^{\dagger }c_{0\sigma  }}
+\tau _{c} \sum_{j\neq 0,\sigma }t^{c}_{0j}\left( c_{0\sigma }^{\dagger }c_{j\sigma }+c_{j\sigma }^{\dagger }c_{0\sigma }\right) \nonumber \\ 
&&+\sum_{\sigma}(\varepsilon _{\sigma }^{d{\rm I}}-\varepsilon _{\sigma }^{d{\rm h}})d_{0\sigma }^{\dagger }d_{0\sigma  }
+\tau _{d} \sum_{j\neq 0,\sigma }t^{d}_{0j}\left( d_{0\sigma }^{\dagger }d_{j\sigma }+d_{j\sigma }^{\dagger }d_{0\sigma }\right) \nonumber \\
&&+(U^{I}-U^{R})n^{d}_{0\downarrow }n^{d}_{0\uparrow },
\end{eqnarray}
$\varepsilon _{\sigma }^{c {\rm I}}$ and $\varepsilon _{\sigma }^{d {\rm I}}$ are the s-p and d impurity state energy levels. The parameters $\tau_{c} $ and $\tau_{d} $  take into
account the change in the hopping energy associated with the
presence of the impurity.\cite{alo2,Oliveira95} 

The last term of Eq.~(\ref{eq:ham1}),
\begin{equation}
H_{M}=-\sum_{l\neq 0,\sigma }\sigma J^{{\rm sd}}\left\langle S^{\rm Co}\right\rangle c_{l\sigma }^{\dagger }c_{l\sigma }
-\sum_{l\neq 0,\sigma }\sigma J^{{\rm dd}}\left\langle S^{\rm Co}\right\rangle d_{l\sigma }^{\dagger }d_{l\sigma },
\end{equation}
is the interaction energy between the magnetic field from the Co ions and the impurity spin. $J^{{\rm sd}}$ and $J^{{\rm dd}}$ are the exchange parameters  and $\left\langle S^{\rm Co}\right\rangle $ is the average magnetic moment at Co sites.

The Green function method allows the calculation of the local density of states and thus the occupation number $n_{0,\sigma}$ for each spin-$\sigma $ orientation. 

From this~\cite{alo2} we find 
\begin{equation}
\widetilde{m}^{\gamma }_{R}(0)=-\frac{1}{\pi }\sum_{\sigma }\int_{-\infty
}^{\epsilon _{{\rm F}}}{\rm Im\,}\,\frac{\sigma\; g^{\gamma }_{00\sigma }(z)}{\alpha_{\gamma } ^{2}-g^{\gamma }_{00\sigma }(z)V_{\rm eff}^{\gamma  \sigma }(z)}\,{\rm d}z  \label{eq:momR}
\end{equation}
for the contribution from rare-earth ions, where $g_{ij\sigma}^{\gamma }(z)$ is the host Green function, $\alpha_{\gamma }=\left( \tau _{\gamma}+1 \right)$ and 
\begin{eqnarray}
\widetilde{m}^{\gamma }_{{\rm ind}}(0)&=&+\frac{1}{\pi }\sum_{\sigma
}\int_{-\infty }^{\epsilon _{{\rm F}}}{\rm Im\,}\frac{\alpha_{\gamma} ^{2}Z_{\rm nn}J^{{\rm \gamma d}}\left\langle S^{\rm Co}\right\rangle }{\left[ \alpha_{\gamma }
^{2}-g^{\gamma }_{00\sigma }(z)\;V_{\rm eff}^{\gamma \sigma
}(z)\right] ^{2}} \nonumber \\ &&\times \left[ \frac{\partial g^{\gamma }_{00\sigma
}(z)}{\partial z}+\left( g^{\gamma }_{00\sigma }(z)\right) ^{2}\right] \;{\rm d}z
\end{eqnarray}
for the contribution from the Co nearest neighbor ions, $\gamma $ referring to the s-p or to the d band. $Z_{\rm nn}$ is the number of Co nearest neighbors. The effective potentials are 
\begin{equation}
V_{\rm eff}^{\gamma \sigma } = V^{\gamma}_{0\sigma}+(\alpha_{\gamma }^{2}-1)(z-\varepsilon^{\gamma {\rm h}}_{\sigma } ),
\end{equation}
where $V^{c}_{0\sigma} = (\varepsilon _{\sigma }^{cI}-\varepsilon _{\sigma }^{c{\rm h}})$ and 
$V^{d}_{0\sigma} = (\varepsilon _{\sigma }^{dI}-\varepsilon _{\sigma }^{d{\rm h}})+(U^{I}-U^{R}) \left\langle n_{-\sigma }^{d}\right\rangle $ are local term potentials. 

Although the impurity \textit{f} electrons were not included in the Hamiltonian (\ref{eq:ham1}), their contribution to the magnetic moment can be found once the impurity \textit{4f} and the rare-earth d polarizations are parallel,\cite{leal87} 
\begin{equation}
\widetilde{m}^{f}(0)= U_{df}\rho_{f}(\varepsilon_F)\widetilde{m}^{d}_{R}(0),
\label{eq:mm}
\end{equation}
where $U_{df}$ is an effective \textit{d-f} Coulomb correlation involving \textit{d} and \textit{f} impurity electrons and $\rho_f(\varepsilon_F)$ is the density of states of the $f$ resonance at the Fermi level energy. 

The magnetic hyperfine field $B_{hf}$ at the impurity site is 
\begin{equation}
B_{hf}=A(Z_{{\rm imp}})\widetilde{m}^{c}(0)+A_{\rm cp}^{5d}\widetilde{m}^{d}(0)+A_{\rm cp}^{4f}\widetilde{m}^{f}(0).
\label{chf}
\end{equation}

\begin{figure}[htbp]
		\includegraphics[width=0.5\textwidth]{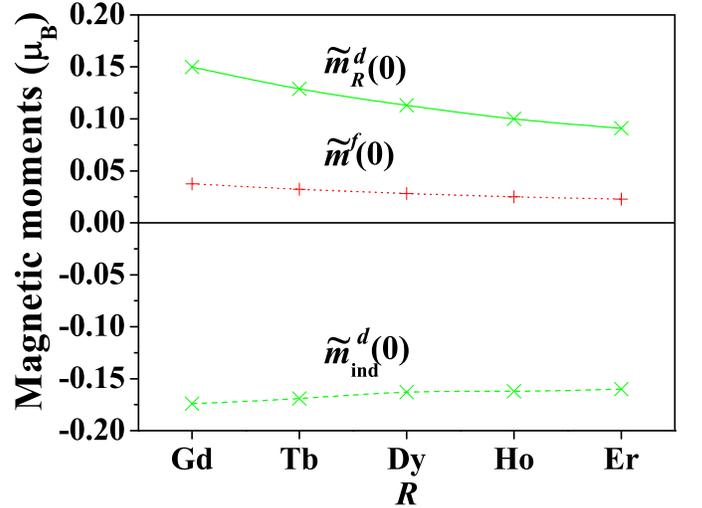}
\caption{(Color online) $d$ and $f$ contributions to the magnetic moment at a Ce impurity diluted in $R$Co$_{2}$. The dotted line corresponds to $\widetilde{m}^{f}(0)$, the solid line to $\widetilde{m}^{d}_R(0)$, and the dashed line to $\widetilde{m}^{d}_{\rm ind}(0)$.}
\label{fig:df}
\end{figure}

In Eq.(\ref{chf}): 
\begin{itemize}
  \item[(i)] $A(Z_{{\rm imp}})$ is the positive Fermi-Segr\`{e} contact coupling parameter. A table of coefficients $A(Z_{{\rm imp}})$ is given by Campbell~\cite{ia} and interpolating linearly between the La and Lu coefficients, one finds for Ce the value $A(Z_{{\rm imp}}) = 322.1 {\rm T}\mu_{B}^{-1}$;
  \item[(ii)]  $A_{\rm cp}^{5d}$  is a negative core polarization  constant of the order of $-120.0{\rm T}\mu_{B}^{-1}$ for the \textit{5d} series;\cite{ia}
  \item[(iii)]  $A_{\rm cp}^{4f}$ is the extra \textit{f}-core negative polarization  constant, the value adopted here being $-150.0{\rm T}\mu_{B}^{-1}$.\cite{leal87}
\end{itemize} 
In Fig.\ref{fig:df}, one sees that the transfered \textit{f}-local magnetization is parallel to $\widetilde{m}^{d}_{R}(0)$ as predicted by Eq.(\ref{eq:mm}). 

\begin{figure}[htbp]
		\includegraphics[width=0.5\textwidth]{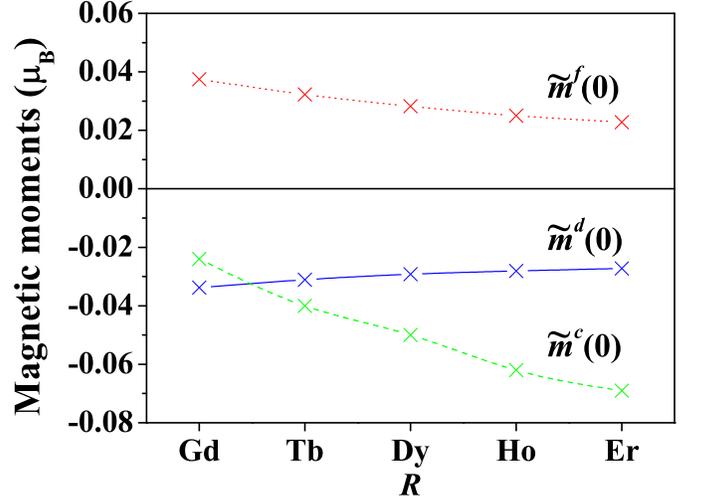}
\caption{(Color online) Magnetic moment contributions at a Ce impurity diluted in $R$Co$_{2}$. The dashed line corresponds to  $\widetilde{m}^{f}(0)$, the dotted line to the total \textit{s-p} $\widetilde{m}^{c}(0)$, and the solid line to the total d-contribution $\widetilde{m}^{d}(0)$.}
\label{fig:mmceco2}
\end{figure}

Fig.~\ref{fig:mmceco2} exhibits the magnetic moments contribution at a Ce impurity diluted in $R$Co$_{2}$: The total
\textit{s-p} magnetic moment, $\widetilde{m}^{c}(0) = \widetilde{m}_{R}^{c}(0)+\widetilde{m}_{\rm ind}^{c}(0)$ , the total \textit{d} contribution  $\widetilde{m}^{d}(0) = \widetilde{m}_{R}^{d}(0)+\widetilde{m}_{\rm ind}^{d}(0)$ and the $\widetilde{m}^{f}(0)$. 

\begin{figure}[htbp]
		\includegraphics[width=0.5\textwidth]{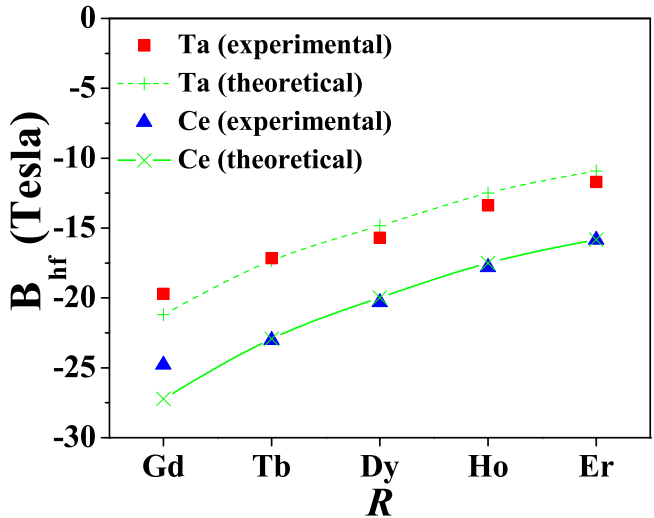}
\caption{ (Color online) Magnetic hyperfine fields for Ce and Ta impurities in $R$Co$_2$.  Triangles (squares) are for Ce (Ta) experimental values.  Full (open) circles represent the theoretical results for Ce (Ta) impurity.}
\label{bhf}
\end{figure}

In Fig.~\ref{bhf} is shown the calculated total magnetic hyperfine fields $B_{hf}$ in comparison with respective experimental data.\cite{saito,note} 
The variation of the magnetic hyperfine field along the series is smooth, its absolute value decreasing as we go from heavy rare-earth GdCo$_2$ to ErCo$_2$.  

Notice that the \textit{4f}-contribution given by Eq.(\ref{eq:mm}) is responsible for the ``shift'' between Ce and Ta hyperfine fields. In both case, we adopted the value $1\mu_{B}$ for the Co magnetic moment in the $R$Co$_{2}$ systems.~\cite{presa}  We stress again that the intermediate valence Ce impurity gives an extra contribution to the core polarization hyperfine fields (see Eqs.(\ref{eq:mm}) and (\ref{chf})), as compared to the \textit{5d} Ta impurity. In this sense Ce impurity acts as a \textit{5d} impurity plus the \textit{4f} resonance contribution to the core polarization. 

\begin{figure}[htbp]\includegraphics[width=0.50\textwidth]{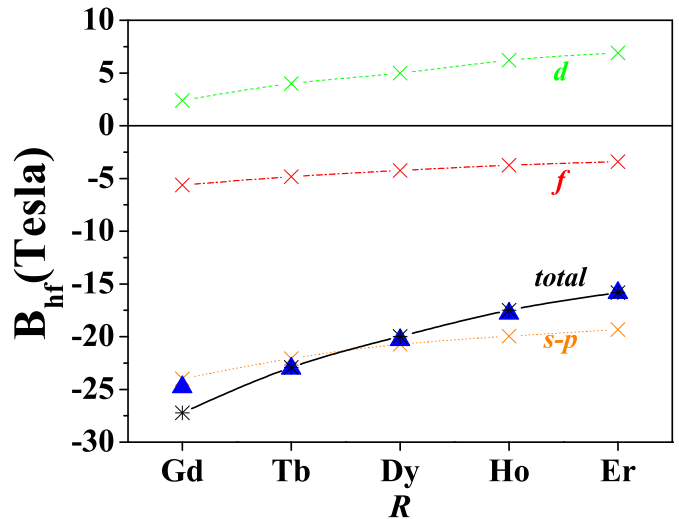}
\caption{ (Color online) Terms in the expression of the magnetic hyperfine fields for Ce  impurities in $R$Co$_2$ according to Eq.(\ref{chf}): from \textit{s-p}, from \textit{d}, from \textit{f} and total.}
\label{bhf2}
\end{figure}

In Fig.(\ref{bhf2}) we display the contributions to the hyperfine field\cite{note} according to  Eq.(\ref{chf}). The \textit{f} term has the value to confirm the difference between Ce and Ta hyperfine fields, as commented above(cf. Fig \ref{bhf}).

Some comments on possible orbital contribution are in order: It is known that most of rare earth impurities exhibits a strong orbital contribution when diluted in Fe, Co and Ni hosts. A theoretical study(\cite{ANT}) uses an Anderson-like model in which the degenerate \textit{4f} energy level of the rare earth impurity is strongly hybridized with a spin polarized electron band. This self-consistent calculation has shown that, for Ce impurity, there is no orbital contribution at all. Here, the host is $R$Co$_2$ and, as mentioned before, since Ce is in an intermediate valence regime, the Ce \textit{4f} impurity level lies very close to the Fermi level of the system. Thus the intermediate valence of Ce results from the competition between the atomic Hund's rule and metallic host effects which destroys a possible f-orbital moment and therefore no such orbital f-moment contribution is expected to appear here either.

\break
\begin{acknowledgments}
We acknowledge support from the Brazilian agencies PCI/MCT and CNPq and useful discussion with Prof.M. Forker and Prof. H. Saitovich.
\end{acknowledgments}


\begin{thebibliography}{00}
\bibitem{note} Experimentally only the absolute value of the field is measured via PAC. In Fig. 3 and 4 we used for the sign the one determined by the theoretical calculation.
\end{thebibliography}
\end{document}